\documentclass[
  aps,
  prl,
  reprint,
  amsmath,
  amssymb,
  superscriptaddress,
  nofootinbib,
  floatfix
]{revtex4-2}

\usepackage{graphicx}
\usepackage{bm}
\usepackage{booktabs}
\usepackage{microtype}
\usepackage{hyperref}

\hypersetup{
  colorlinks=true,
  linkcolor=blue,
  citecolor=blue,
  urlcolor=blue
}

\hypersetup{pdftitle={Binary Pulsars as a Three-Dimensional Solar System Accelerometer}}

\newcommand{\Msun}{M_\odot}
\newcommand{\uasyr}{\mu\mathrm{as}\,\mathrm{yr}^{-1}}

\newcommand{\xzqc}[1]{{\textcolor{magenta} {-[XZQ Comment: #1]-}}}

\usepackage{xcolor}
\usepackage{comment}

\begin{document}


\title{Binary‑pulsar Timing: A 3D Solar‑System Accelerometer and Unknown‑Source Probe}

\author{Zheng-Long Wang}
\affiliation{Key Laboratory of Dark Matter and Space Astronomy, Purple Mountain Observatory, Chinese Academy of Sciences, Nanjing 210033, People's Republic of China}
\affiliation{School of Astronomy and Space Science, University of Science and Technology of China, Hefei, Anhui 230026, People's Republic of China}

\author{Zi-Qing Xia}
\email{xiazq@pmo.ac.cn}
\affiliation{Key Laboratory of Dark Matter and Space Astronomy, Purple Mountain Observatory, Chinese Academy of Sciences, Nanjing 210033, People's Republic of China}

\author{Bo Zhang}
\affiliation{Key Laboratory of Dark Matter and Space Astronomy, Purple Mountain Observatory, Chinese Academy of Sciences, Nanjing 210033, People's Republic of China}
\affiliation{School of Astronomy and Space Science, University of Science and Technology of China, Hefei, Anhui 230026, People's Republic of China}

\author{Yi-Zhong Fan}
\email{yzfan@pmo.ac.cn}
\affiliation{Key Laboratory of Dark Matter and Space Astronomy, Purple Mountain Observatory, Chinese Academy of Sciences, Nanjing 210033, People's Republic of China}
\affiliation{School of Astronomy and Space Science, University of Science and Technology of China, Hefei, Anhui 230026, People's Republic of China}

\date{\today}

\begin{abstract}
The acceleration of the Solar System barycenter (SSB) offers a unique precision probe of the local gravitational environment, enabling searches for otherwise invisible gravitating sources such as Planet Nine and nearby (primordial) black holes. Using the binary-pulsar timing measurements, we develop an analysis framework that combines the line-of-sight differential accelerations of 26 binary pulsars to jointly reconstruct the three-dimensional acceleration of the SSB and place directional upper limits. Across three smooth Galactic-potential baselines, the residual acceleration is consistent with zero. We obtain directional 95\% upper limits of $0.198$, $0.262$, and $0.693~\mu\mathrm{as}\,\mathrm{yr}^{-1}$ over 50\%, 75\%, and all sampled directions, respectively. These limits are tighter by factors of 3.5, 3.6, and 1.6 than a matched Gaia EDR3 benchmark and improve previous pulsar constraints by more than an order of magnitude. We further extend the framework to the full SSB--pulsar two-endpoint response, enabling direct position-dependent mass constraints on unknown gravitational sources. A $10\,M_\odot$ object is excluded within 0.39, 0.34, and 0.21 pc over the same sky fractions. Our unknown-mass limits surpass the tidal-equivalent INPOP19a reference beyond 0.2 pc by about an order of magnitude over most of the sky at 1 pc.
\end{abstract}

\maketitle

\textit{Introduction.--}
Precision measurements of the Solar System barycenter (SSB) acceleration can probe the local gravitational field and reveal gravitating matter independently of luminosity. 
Candidate perturbers include unknown Solar-System bodies \cite{Trujillo2014,Batygin2016,Batygin2019PNreview, BrownBatygin2021,Guo2018,Caballero2018} (such as Planet Nine), and stellar-mass black holes \cite{Elbert2018,LIGO2023Population,ElBadry2023BH1}, wandering intermediate-mass black holes \cite{Greene2020,Weller2022}, primordial black holes \cite{Carr2021,SetoCooray2007,Dror2019}, and even compact dark structures \cite{RicottiGould2009,BringmannScottAkrami2012,Urrutia2023}.
Planetary ephemerides provide complementary local-gravity constraints
\cite{PitjevPitjeva2013,Fienga2020P9}. The SSB acceleration is also an
observer-frame quantity: its transverse projection produces the
secular-aberration glide measured with VLBI and Gaia quasars
\cite{Titov2011,MacMillan2019,GaiaEDR3Acceleration}, while TESS offers a
complementary stellar-astrometry route
\cite{Gai2022TESSAstrometry,Daniel2026TESSAberration}. Binary-pulsar orbital
clocks provide an independent dynamical probe of the same observer motion.

Galactic differential accelerations have long entered precision binary-pulsar orbital-decay measurements \cite{DamourTaylor1991}, and pulsar timing now probes the Galactic acceleration field directly \cite{Phillips2021,Chakrabarti2021,Moran2024,Donlon2024,Donlon2025}. 
For a
detached binary, the orbital-period derivative measures the line-of-sight
acceleration relative to the SSB after intrinsic and kinematic terms are
removed, without the magnetic-braking model required for isolated-pulsar
spin-down estimates \cite{LorimerKramer2005}. Individual SSB--pulsar
acceleration constraints already exist \cite{Deller2008}; each binary supplies
one line-of-sight projection, so a network is required to reconstruct the common three-dimensional mode. 
This quasi-static orbital-clock channel complements time-dependent flyby searches and Solar-System ephemeris or orbital-perturbation tests
\cite{SetoCooray2007,Dror2019,Champion2010,Caballero2018}.
Zakamska and Tremaine \cite{ZakamskaTremaine2005} 
pioneered the use of precision astronomical clocks, 
including binary-pulsar orbital clocks, 
to constrain a peculiar Solar-System acceleration; subsequent timing and 
VLBI astrometry of PSR J0437$-$4715 provided a tighter single-system bound
\cite{Deller2008}. 
Related pulsar-acceleration measurements have also been used to
probe dark substructure and satellite-induced Galactic disequilibrium 
\cite{Chakrabarti2026,Wang2026,Donlon2026}.

Here we combine the line-of-sight differential accelerations of 
26 binary pulsars to reconstruct the common three-dimensional SSB acceleration 
vector, together with its full covariance and joint significance, and compare 
the resulting all-sky limits with quasar astrometry. 
Beyond the common SSB acceleration, we further retain the 
pulsar-side response to a local perturber, yielding the full two-endpoint 
SSB--pulsar gravitational response and thereby position-dependent constraints 
on unknown gravitational sources.

\begin{figure*}[t]
  \centering
  \includegraphics[width=0.98\textwidth]{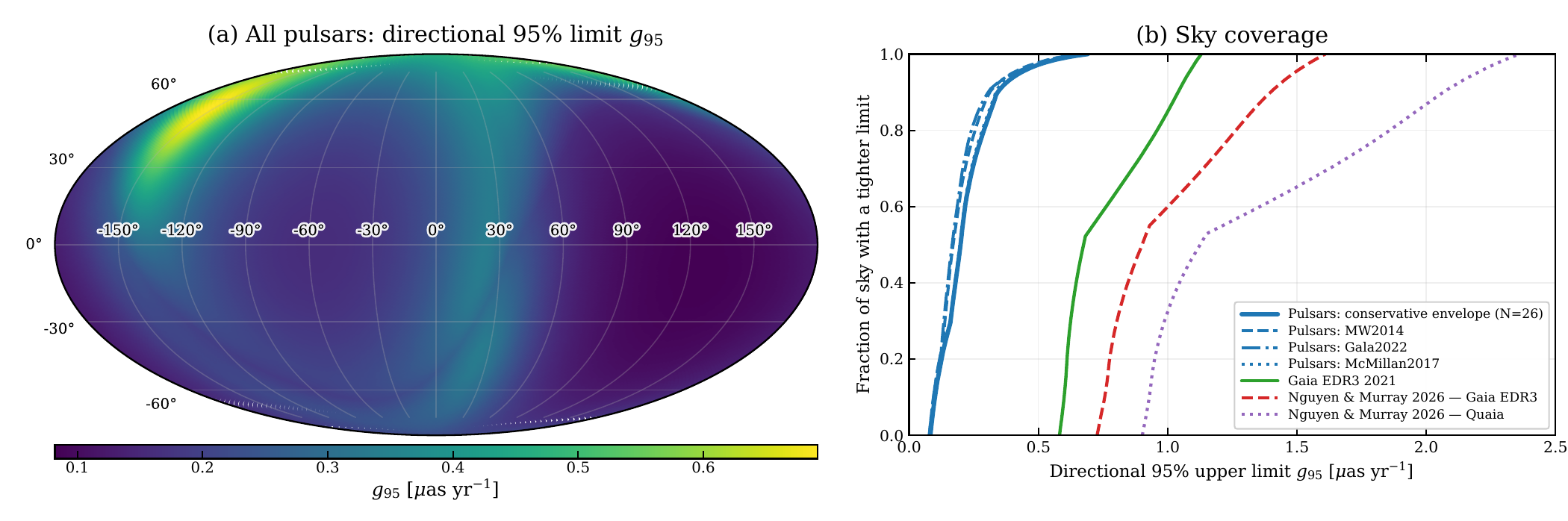}
  \caption{\textbf{All-sky additional-acceleration limits.}
  (a) Directional 95\% upper limits $g_{95}$ from the 26-binary pulsar sample,
  showing the conservative envelope over the three Galactic-potential
  baselines; the map coordinate is the trial acceleration apex.
  (b) Sky fraction with a tighter limit. The blue curves show the
  baseline-specific pulsar limits: Gala2022 (solid), MW2014 (dashed),
  and McMillan2017 (dotted). The green solid curve is the conservative matched Gaia EDR3 2021 benchmark; the red dashed and purple dotted curves
  show the approximate systematics-aware Gaia-EDR3 and Quaia benchmarks,
  respectively \cite{GaiaEDR3Acceleration,NguyenMurray2026}.}
  \label{fig:accel}
\end{figure*}

\textit{Data and analysis.--}
We use the 26 binary pulsars in the Donlon \textit{et al.} final-v3 compilation \cite{Donlon2025} that have the timing and astrometric inputs needed to evaluate Eqs.~(\ref{eq:atim})--(\ref{eq:agal}), including the orbital period and its derivative, proper motion, distance, sky position, and the adopted GR orbital-decay correction. Final-v3 inputs are retained by default; updated published inputs are adopted only for the seven systems listed in the End Matter. The update choice is independent of the acceleration residual.

For binary $i$, the timing-derived line-of-sight differential acceleration relative to the SSB is
\begin{equation}
 a^{\rm tim}_{i}=\frac{c}{P_{b,i}}
 \left(\dot P^{\rm obs}_{b,i}-\dot P^{\rm GW}_{b,i}
 -\frac{P_{b,i}\mu_i^2d_i}{c}\right),
 \label{eq:atim}
\end{equation}
where $P_b$ is the orbital period, $\dot P_b^{\rm obs}$ the observed orbital period derivative, and $\dot P_b^{\rm GW}$ the intrinsic orbital decay from gravitational-wave damping in general relativity \cite{Peters1964,DamourTaylor1992}. 
This is the GR correction tabulated for the adopted detached-binary acceleration sample, and its quoted uncertainty is propagated in the Monte Carlo. Here $\mu$ is the total proper motion and $d$ the distance; the final term is the Shklovskii contribution \cite{Shklovskii1970,BellBailes1996}. 
We propagate the quoted timing and astrometric uncertainties with $10^6$ joint Monte Carlo realizations.
Following the realization-by-realization parallax treatment of Wang \textit{et al.} \cite{Wang2026}, we sample the parallax $\varpi$ from its quoted Gaussian measurement conditioned on $\varpi>0$ and convert each draw through $\varpi=1/d$(where $d$ is the distance of pulsar); the same draw enters the Shklovskii and Galactic terms. 
Because low-significance parallaxes generate extended distance tails under this
inversion, we repeat the vector analysis with
$\varpi/\sigma_\varpi>3$ and $>5$ cuts and find negligible changes (see the End Matter for details). 
Full timing-solution covariance products are not uniformly available for this heterogeneous literature sample; we therefore use the published scalar timing and astrometric uncertainties independently within each binary as a homogeneous baseline.

We repeat the analysis independently for three Galactic potential models: MW2014 \cite{Bovy2015} (hereafter MW2014), Gala MilkyWayPotential2022 \cite{PriceWhelan2017,PriceWhelan2022MWP} (hereafter Gala2022), and the McMillan2017 posterior \cite{McMillan2017}. 

For each Galactic-potential model, the predicted line-of-sight
differential acceleration of binary $i$ is
\begin{equation}
 a_i^{\rm Gal}
 =
 \left[
 \bm a_{\rm Gal}(\bm x_i)-\bm a_{\rm Gal}(\bm x_\odot)
 \right]\cdot\hat{\bm n}_i ,
 \label{eq:agal}
\end{equation}
where $\bm a_{\rm Gal}(\bm x)$ is the acceleration field,
$\bm x_i$ and $\bm x_\odot$ are the pulsar and Solar-System barycenter
(SSB) positions, and $\hat{\bm n}_i$ points from the SSB to the pulsar.
We adopt $(R_0,z_\odot)=(8.178~{\rm kpc},20.8~{\rm pc})$, corresponding to
$\bm x_\odot=(-R_0,0,z_\odot)$ in our Galactocentric Cartesian convention
\cite{Gravity2019R0,BennettBovy2019}.

The three potentials yield similar smooth background predictions for our sample.
Only 7/26, 6/26, and 7/26 binaries, respectively, have a modeled
SSB--pulsar differential Galactic acceleration larger than the
Monte Carlo standard deviation of their timing-derived acceleration.
We therefore adopt these published smooth potentials as our baselines and constrain a coherent excess acceleration, rather than jointly fitting a flexible Galactic field and a free observer dipole. 
Source-level diagnostics and the scope of this treatment are given in the End Matter.


For the Monte Carlo realization $k$, we define $R_i^{(k)}=a_i^{{\rm tim},(k)}-a_i^{{\rm Gal},(k)}$ and, for $N=26$
binaries, $\bm R^{(k)}=(R_1^{(k)},\ldots,R_N^{(k)})^T$. The residual covariance is estimated as $\bm C_R={\rm Cov}_{\rm MC}[\bm R^{(k)}]$.

For MW2014 and Gala2022, the off-diagonal elements of $\bm C_R$ are set to zero. 
For McMillan2017, all pulsars share the same potential realization in each joint draw, so that the draws retain the model-induced cross-source covariance.
The resulting weight matrix $\bm W=\bm C_R^{-1}$ is held fixed across Monte Carlo realizations.

We first investigate the common acceleration of the SSB, considering only acceleration perturbations in the vicinity of the Solar system.
The pulsar-side perturbation is subsequently incorporated via the full SSB--pulsar two-endpoint response. 
For binary $i$, the residual can be modeled as
\begin{equation}
 R_i = -\hat{\bm n}_i\cdot\bm a_{\rm loc} +\delta_i^{\rm Gal} +\epsilon_i ,
 \label{eq:vector_model}
\end{equation}
where $\bm a_{\rm loc}$ denotes an additional coherent SSB acceleration, $\delta_i^{\rm Gal}$ denotes remaining Galactic-model mismatch and $\epsilon_i$ measurement noise. Equation~(\ref{eq:vector_model}) describes the common SSB-side response. 
We do not introduce an unrestricted nuisance model for $\delta_i^{\rm Gal}$, but instead carry the three Galactic baselines separately.

We use Galactic Cartesian axes with $+X$ toward the Galactic center
$(l=0^\circ,b=0^\circ)$, $+Y$ toward $l=90^\circ$, and $+Z$ toward the
North Galactic Pole. Defining $H_{i\alpha}=-\hat n_{i,\alpha}$ for
$\alpha\in\{X,Y,Z\}$, the reconstructed three-dimensional acceleration
$\bm a_{\rm rec}^{(k)}$ of $\bm a_{\rm loc}$ in realization $k$ is
\begin{equation}
 \bm a_{\rm rec}^{(k)}
 =
 \left(\bm H^T\bm W\bm H\right)^{-1}
 \bm H^T\bm W\,\bm R^{(k)} .
 \label{eq:avec}
\end{equation}
We characterize the ensemble by
$\bar{\bm a}_{\rm rec}=\langle\bm a_{\rm rec}^{(k)}\rangle_{\rm MC}$ and
$\bm C_a={\rm Cov}_{\rm MC}[\bm a_{\rm rec}^{(k)}]$.
To quantify the consistency of the reconstructed acceleration with zero,
we define the quadratic-form statistic
$q=\bar{\bm a}_{\rm rec}^{\,T}\bm C_a^{-1}\bar{\bm a}_{\rm rec}$.
Under the null hypothesis of zero residual acceleration, $q$ follows a
$\chi^2$ distribution with three degrees of freedom corresponding to the
three Cartesian components of $\bm a_{\rm rec}$.
The corresponding tail probability is
$p_{3{\rm D}}^{\rm QF}=P(\chi^2_3\ge q)$.
For the three potential ensembles, a centered Monte Carlo calibration agrees
with this quadratic-form probability to better than $3\times10^{-4}$.

To quantify the sensitivity to an additional acceleration along an arbitrary
direction, we project the residual vector onto a trial direction
$\hat{\bm\Omega}$. We write
$\bm a_{\rm loc}=A\hat{\bm\Omega}$, where $A$ is the
acceleration amplitude. Defining the design matrix
$H_{i\alpha}=-\hat n_{i,\alpha}$ for
$\alpha\in\{X,Y,Z\}$, the corresponding response vector is
$\bm s=\bm H\hat{\bm\Omega}$. The signed one-parameter
Generalized Least-Squares (GLS) estimator of $A$ in realization $k$ is
\begin{equation}
 A^{(k)}(\hat{\bm\Omega})
 =
 \frac{\bm s^T\bm W\,\bm R^{(k)}}
      {\bm s^T\bm W\,\bm s}.
 \label{eq:aproj}
\end{equation}
For each Galactic-potential baseline $\mathcal M$, let
$A^{(k)}(\hat{\bm\Omega}\mid\mathcal M)$ denote the signed GLS estimator
of Eq.~(5) in realization $k$. We define its Monte Carlo mean
$\bar A\equiv\langle A^{(k)}\rangle_k$ and empirical one-sided width
$w_{95}\equiv-Q_{0.05}[A^{(k)}-\bar A]$, where $Q_{0.05}$ denotes the
empirical 5th percentile of the centered Monte Carlo samples. The minus
sign converts the lower-tail excursion into a positive one-sided width.
The directional 95\% upper limit is
\begin{equation}
\begin{split}
A_{95}(\hat{\bm\Omega}\mid\mathcal M)
&=\max[0,\bar A(\hat{\bm\Omega}\mid\mathcal M)]+w_{95},\\
A_{95}(\hat{\bm\Omega})
&=\max_{\mathcal M}A_{95}(\hat{\bm\Omega}\mid\mathcal M).
\end{split}
\label{eq:a95}
\end{equation}
No Gaussian form is assumed. The first maximum imposes $A\geq0$, while
the second gives the conservative envelope over the three Galactic
baselines. We quote $g_{95}\equiv A_{95}/c$.

\textit{Common-acceleration result.--}
We reconstruct the additional SSB acceleration common to the pulsar samples using the orbital-period derivatives of 26 binary pulsars.
We express this acceleration in acceleration-equivalent
angular-rate units as
$\bm g_{\rm loc}\equiv\bm a_{\rm loc}/c$.
For all three smooth Galactic baselines, the reconstructed vectors remain consistent
with zero in the joint three-dimensional summary (Table~\ref{tab:vector}):
$p_{3{\rm D}}^{\rm QF}=0.216$, 0.197, and 0.201 for MW2014,
Gala2022, and McMillan2017. 
The largest component offset is the positive $g_Z$, at only
$1.5$--$1.9\sigma$ across the three baselines.
The results are stable across the three Galactic-potential baselines.
The resulting conservative directional upper-limit map and
baseline-specific sky-coverage curves are shown in
Fig.~\ref{fig:accel}.
We therefore use the conservative directional upper-limit map
as the primary summary.

\begin{table}[t!]
\caption{\label{tab:vector}
Three-dimensional differential acceleration vector in
acceleration-equivalent angular-rate units.
The components $g_X$, $g_Y$, and $g_Z$ are the projections of
$g_{\rm loc}=a_{\rm loc}/c$ onto the Galactic Cartesian axes.
Components are in $\uasyr$ in the Galactic Cartesian frame:
$+X$ toward the Galactic center, $+Y$ toward $l=90^\circ$, and
$+Z$ toward the North Galactic Pole.
Quoted uncertainties are Monte Carlo component standard deviations;
$p_{3{\rm D}}^{\rm QF}$ follows the definition given in the text.}
\begingroup
\scriptsize
\setlength{\tabcolsep}{2.8pt}
\begin{ruledtabular}
\begin{tabular}{@{}lcccc@{}}
Model & $g_X$ & $g_Y$ & $g_Z$ & $p_{3\rm D}^{\rm QF}$\\
\hline
MW2014 & $-0.072\!\pm\!0.131$ & $+0.060\!\pm\!0.092$ & $+0.300\!\pm\!0.165$ & 0.216\\
Gala2022 & $-0.152\!\pm\!0.129$ & $+0.110\!\pm\!0.091$ & $+0.313\!\pm\!0.165$ & 0.197\\
McMillan2017 & $-0.058\!\pm\!0.147$ & $+0.086\!\pm\!0.096$ & $+0.254\!\pm\!0.167$ & 0.201\\
\end{tabular}
\end{ruledtabular}
\endgroup
\end{table}

For the matched Gaia EDR3 2021 benchmark, we use the published Galactic
Cartesian glide vector
$\bm g_{\rm obs}=(5.04,-0.10,-0.29)~\uasyr$
and its full $3\times3$ covariance
\cite{GaiaEDR3Acceleration}. For each Galactic-potential baseline
$\mathcal M$, we form
$\bm r_{\mathcal M}^{\rm Gaia}
=\bm g_{\rm obs}
-\bm g_{{\rm MW},\mathcal M}(\bm x_\odot)
-\bm g_{\rm LMC}$,
where the Milky-Way term is the \emph{absolute} SSB acceleration rather
than the SSB--pulsar differential acceleration. For a trial direction
$\hat{\bm\Omega}$, we use
$\widehat A_{\rm Gaia}
=(\hat{\bm\Omega}^{T}\bm C^{-1}\bm r_{\mathcal M}^{\rm Gaia})/
(\hat{\bm\Omega}^{T}\bm C^{-1}\hat{\bm\Omega})$
and
$\sigma_{A,{\rm Gaia}}
=(\hat{\bm\Omega}^{T}\bm C^{-1}\hat{\bm\Omega})^{-1/2}$.
The matched one-sided 95\% Gaussian limit is
$A_{95}^{\rm Gaia}
=\max(0,\widehat A_{\rm Gaia})
+1.64485\,\sigma_{A,{\rm Gaia}}$,
the Gaussian analogue of the empirical boundary-aware construction
used for the pulsar analysis.
For the McMillan2017 baseline, $\bm C$ also includes the SSB-acceleration
covariance from the same potential posterior used in the pulsar analysis.

\begin{figure*}[t]
  \centering
  \includegraphics[width=0.96\textwidth]{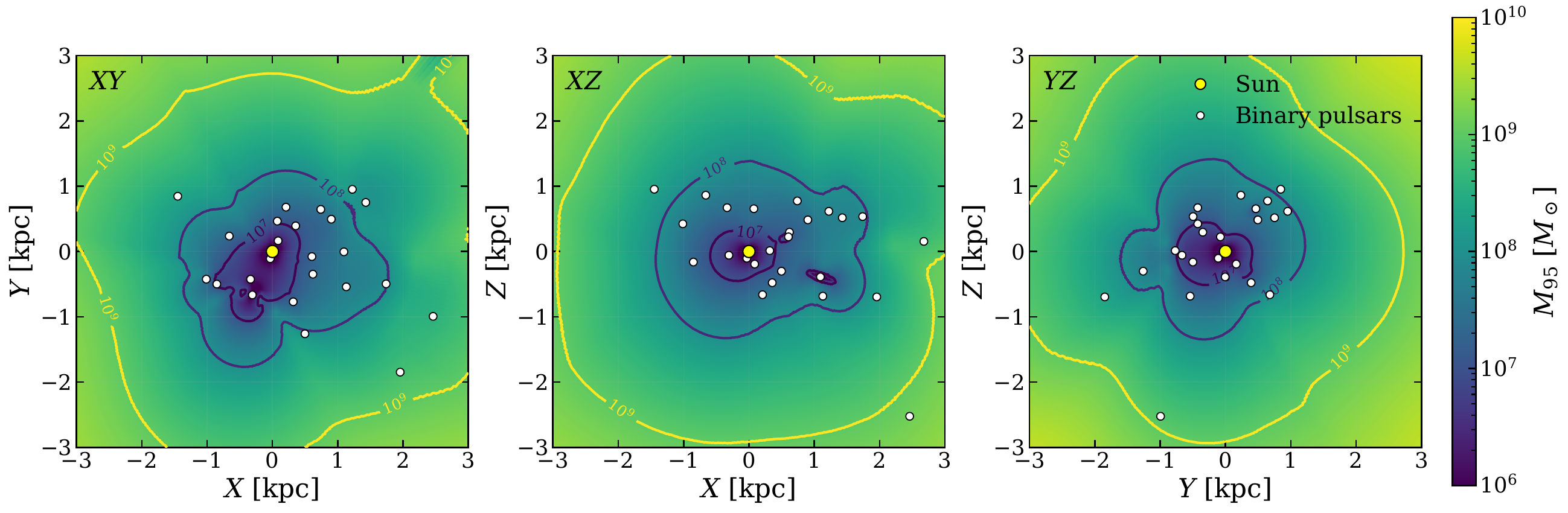}
  \caption{\textbf{Full SSB--pulsar response.}
  Monte-Carlo-propagated $M_{95}$ in three orthogonal heliocentric Galactic
  planes, shown over $\pm3$ kpc using all 26 binaries. White markers show the
  projected pulsar positions in each plane and the yellow marker denotes the
  Sun; contours mark $10^7$, $10^8$, and
  $10^9\,\Msun$. The plotted field interpolates the directional all-sky scan
  and is used to display the signed two-endpoint SSB--pulsar differential structure;
  headline numerical values are taken from the underlying scan rather than
  read from this image.}
  \label{fig:spatial}
\end{figure*}

The Large Magellanic Cloud (LMC) affects the two observables differently.
Pulsar timing measures a differential acceleration, so the leading common
LMC pull cancels between the Sun and pulsar. Using the actual 26-source geometry
and the Hernquist LMC models of Erkal \textit{et al.}
\cite{Erkal2019LMC}, the remaining tidal pattern projects onto the reconstructed common vector
with a maximum amplitude of only $0.0020~\uasyr$ over
$M_{\rm LMC}=(1$--$2)\times10^{11}\,\Msun$ and the three Galactic baselines,
about 1\% of the median-sky pulsar limit. We therefore omit the direct LMC
force from the pulsar baseline at the present precision, while retaining it
for the absolute Gaia acceleration. These models are
normalized to
$M(<8.7\,{\rm kpc})=1.7\times10^{10}\,\Msun$
\cite{vanDerMarelKallivayalil2014} at $D_{\rm LMC}=49.59$ kpc
\cite{Pietrzynski2019LMC}, with the mass bracket from
Ref.~\cite{Vasiliev2023LMC}. This estimate concerns the direct LMC force;
satellite-induced Galactic disequilibrium is not separately fitted and remains
part of the Galactic-field mismatch \cite{Donlon2026}. After these foreground
treatments, the pulsar and Gaia curves provide complementary constraints on the same additional observer-frame acceleration.

As shown in the right panel of Fig.~\ref{fig:accel}, our conservative
least-favorable pulsar limits are tighter than the matched Gaia EDR3 2021
curve by factors 3.5, 3.6, and 1.6 at 50\%, 75\%, and 100\% sky coverage,
respectively.
Nguyen and Murray \cite{NguyenMurray2026} marginalize additional angular systematics and obtain wider credible intervals; because a full numerical component covariance is not tabulated, our converted Gaia-EDR3 and Quaia curves are approximate benchmarks. 
Relative to their Gaia-EDR3 curve, the corresponding improvement factors are about 4.6, 4.6, and 2.3. 
For historical context, our limits are about 12--28 times tighter
than the binary-pulsar constraints of Zakamska and Tremaine
\cite{ZakamskaTremaine2005}, and about 13 times tighter than the
50\%-sky bound of Deller et al. \cite{Deller2008}.

\textit{Probing unknown source with the Full SSB--pulsar response.--}
Here we account for the full SSB--pulsar response to search unknown quasi-static sources.
We restrict the point-mass interpretation to quasi-static perturbers whose displacement over the timing baseline is small compared with the relevant source--SSB and source--pulsar separations, so that the induced differential acceleration is approximately constant.
Our work complements searches for time-dependent flyby signatures and
Solar-System orbital perturbations
\cite{SetoCooray2007,Champion2010,Guo2018,Caballero2018,Dror2019}.

For a quasi-static perturber of mass $M$ at heliocentric position $\bm r$, with pulsar $i$ at
$\bm r_i=d_i\hat{\bm n}_i$ (where $d_i$, $\hat{\bm n}_i$ is the distance and direction of pulsar $i$), the induced line-of-sight differential
acceleration is
\begin{equation}
 a_{{\rm pert},i}
 =
 GM\!\left[
 \frac{\bm r-\bm r_i}{|\bm r-\bm r_i|^3}
 -
 \frac{\bm r}{|\bm r|^3}
 \right]\!\cdot\hat{\bm n}_i ,
 \label{eq:response}
\end{equation}
where $G$ is the gravitational constant, and the two terms are the pulsar-side and SSB-side accelerations,
respectively.

At fixed $\bm r$, Eq.~(\ref{eq:response}) is linear in the physical mass
$M\geq0$ and defines a signed template $\bm t(\bm r)$ per unit mass, with
$t_i(\bm r)=a_{{\rm pert},i}/M$. For each Galactic baseline, the
unconstrained estimator $\widehat M$ is obtained with the same one-parameter
GLS projection as Eq.~(\ref{eq:aproj}), replacing $\bm s$ by
$\bm t(\bm r)$. The 95\% upper limit $M_{95}(\bm r)$ is constructed with the same
centered lower-tail inversion as Eq.~(\ref{eq:a95}), and the reported
limit is the conservative envelope over the three Galactic baselines.

\begin{figure}[!t]
  \centering
  \includegraphics[width=0.96\columnwidth]{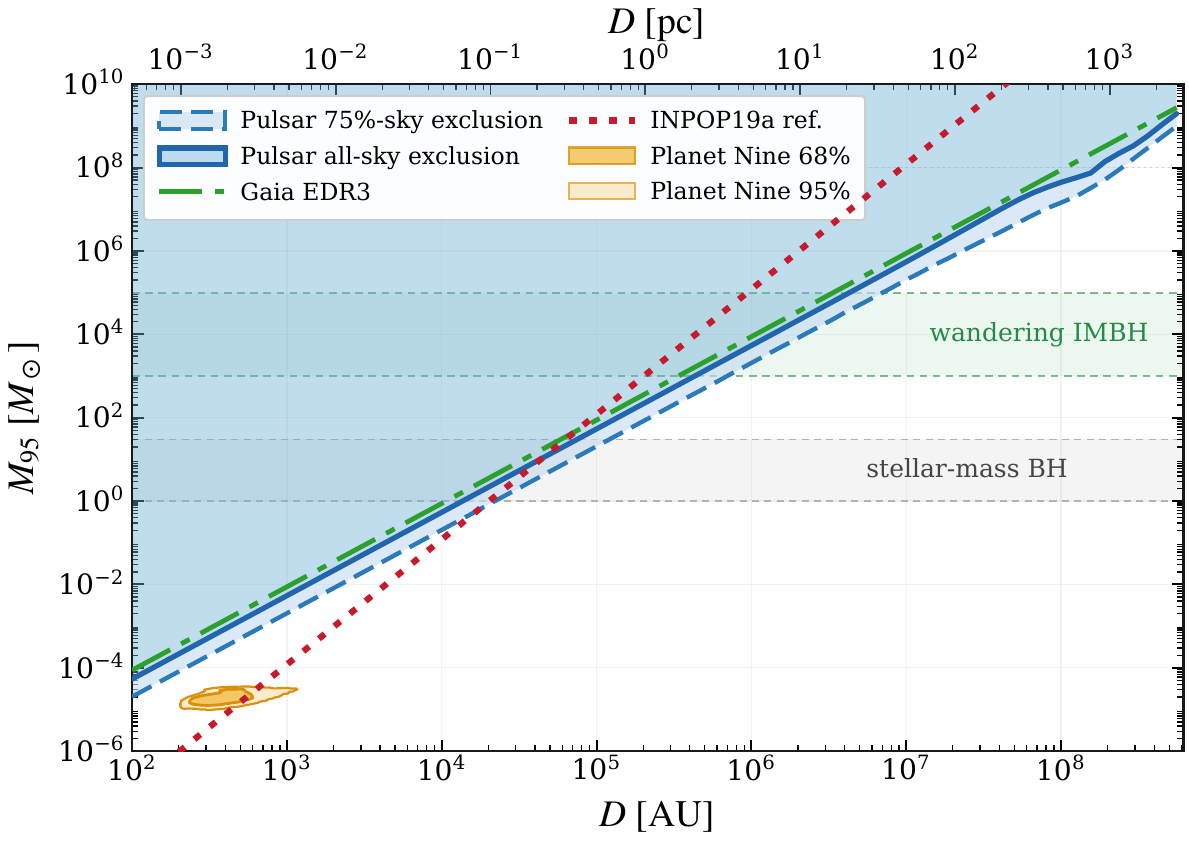}
  \caption{\textbf{Unknown-source limits.}
  Constraints on quasi-static point masses from the 26-binary pulsar sample.
  Blue shaded regions show the excluded parameter space, with lighter and
  darker regions corresponding to 75\%-sky and all-sky coverage, respectively;
  dashed and solid boundaries give the corresponding 95\% upper limits
  $M_{95}(D)$. The green curve shows the Gaia EDR3 2021 observer-side
  acceleration benchmark, and the red dotted curve is the INPOP19a
  tidal-equivalent reference normalized to $5\,M_\oplus$ at 500 AU
  \cite{Fienga2020P9}. Gold regions enclose the 68\% and 95\% Brown--Batygin
  Planet Nine reference populations \cite{BrownBatygin2021}.
  Horizontal bands indicate stellar-mass black holes and wandering IMBHs
  \cite{Elbert2018,LIGO2023Population,Greene2020,Weller2022}.}
  \label{fig:landscape}
\end{figure}

The exact response becomes quantitatively different from an observer-only
$GM/D^2$ (where $D$ is the perturber distance)  extrapolation once $D$ approaches the pulsar distances. At
$D=0.934$ kpc, the full SSB--pulsar calculation gives median-sky and
least-favorable limits of $3.41\times10^7$ and
$1.40\times10^8\,\Msun$, compared with $5.70\times10^7$ and
$2.00\times10^8\,\Msun$ from the observer-only extrapolation. Retaining the
pulsar-side term therefore improves these benchmarks by factors 1.67 and
1.43, demonstrating the additional source-space information carried by the
second endpoint. Fig.~\ref{fig:spatial} also shows the three geometric
regimes of Eq.~(\ref{eq:response}). For $D\ll d_i$, the SSB term dominates and
$M_{95}\propto D^2$. For $D\sim d_i$, the two endpoint terms become comparable,
producing the lobes, valleys, and cancellations in source space. For
$D\gg d_i$, the common acceleration cancels to leading order and the
remaining tidal response is of order $GMd_i/D^3$, so a fixed acceleration
threshold corresponds to $M_{95}\propto D^3$.

Within the quasi-static point-mass model, a $ \sim10\,\Msun$ object is excluded out
to 0.39, 0.34, and 0.21 pc over 50\%, 75\%, and all sampled directions. For
$10^2$, $10^3$, and $10^4\,\Msun$, the least-favorable reaches are 0.66, 2.09,
and 6.61 pc. For comparison, we define an INPOP19a planetary-ephemeris
\emph{tidal-equivalent reference} by normalizing its published constraint to $5\,M_\oplus$ at 500 AU \cite{Fienga2020P9} and scaling as
$M\propto D^3$; this gives 0.46, 0.98, and 2.12 pc at the same masses. This is
a source-space scale comparison, not a confidence-matched exclusion curve.

\textit{Discussion.--}
Fig.~\ref{fig:landscape} shows the 95\% upper‑limit mass \(M_{95}\) for unknown sources as a function of distance D.
The stellar-mass black-hole band spans
$1$--$30\,\Msun$, while the wandering intermediate-mass black-hole band spans
$10^3$--$10^5\,\Msun$
\cite{Elbert2018,LIGO2023Population,Greene2020,Weller2022}. Primordial black
holes and compact minihalos can span broader, model-dependent masses
\cite{Carr2021,RicottiGould2009,BringmannScottAkrami2012,Urrutia2023}, for
which the continuous $M_{95}(D)$ reach is more informative. The INPOP19a curve is the tidal-equivalent reference defined above, so
planetary dynamics and pulsar timing should be read as complementary
source-space benchmarks.
The Planet Nine reference population in Fig.~\ref{fig:landscape}
is not yet constrained by the present pulsar sensitivity, although future
timing and network improvements may begin to test part of this parameter
space independently \cite{BrownBatygin2021}.

Based on timing solutions from 26 binary pulsars, we establish a three-dimensional dynamical probe of the Solar System's local gravitational environment, sensitive to both a common SSB acceleration and the position-dependent response of unknown sources. 
Unknown gravitational sources may still exist within the regions of source space not excluded by our present limits.
Longer timing baselines, improved parallaxes, and a larger high-precision binary sample will extend the sensitivity toward lower masses and larger distances, progressively probing this remaining parameter space. 
A denser pulsar network can ultimately turn binary-pulsar timing into an increasingly sensitive, luminosity-independent probe of the local gravitational environment.

\textit{Acknowledgments.--}This work is supported by the Astrometric Reference Frame project (No. JZZX-020501), the National Key Research and Development Program of China (No. 2022YFF0503304) and  the Strategic Priority Research Program of the Chinese Academy of Sciences (No. XDB0550400). Y.Z.F thanks the support of New Cornerstone Science Foundation through the XPLORER PRIZE. 

\bibliography{references}

\clearpage
\appendix

\section{Sample definition, residual diagnostics, and Galactic-potential scope}
\label{app:data}

\subsection{Adopted sample and parallax robustness}

The headline analysis uses the full 26-source detached, non-accreting binary
sample satisfying the common requirements of the final-v3 compilation
\cite{Donlon2025}. Relative to the final-v3 numerical baseline, updated
published inputs are adopted for seven systems, independently of their
acceleration residuals. J0437$-$4715 uses the precision timing solution of
Reardon \textit{et al.} \cite{Reardon2024}; J0613$-$0200 uses EPTA DR2
timing and astrometry \cite{EPTA2023}; J0737$-$3039A/B uses the 16-yr
double-pulsar timing solution of Kramer \textit{et al.} \cite{Kramer2021};
J1012+5307 uses the EPTA DR2 orbital timing solution \cite{EPTA2023};
J1022+1001 uses EPTA DR2 astrometry \cite{EPTA2023}; J1600$-$3053 combines
the EPTA DR2 orbital timing solution \cite{EPTA2023} with the PPTA DR2
timing parallax of Reardon \textit{et al.} \cite{Reardon2021}; and
J1909$-$3744 uses the 15-yr timing solution of Liu \textit{et al.}
\cite{Liu2020}. All remaining binaries retain the final-v3 inputs.

Given the timing and astrometric precision across the
sample, we test the robustness of the common-vector inference to
low-significance parallaxes and individual-source influence. Requiring the signal‑to‑noise ratio of parallax
$\varpi/\sigma_\varpi>3$ and $>5$ leaves 21 and 17 binaries,
respectively. The $3\sigma$ cut leaves the fitted vectors unchanged at
the quoted precision, while the stricter $5\sigma$ cut shifts no
Cartesian component by more than $0.066$ of its fiducial $1\sigma$
uncertainty. Leave-one-out tests likewise show that no single binary
dominates the inference, with a maximum component shift of
$0.140\,\mu{\rm as\,yr^{-1}}$.

\subsection{Source-level diagnostics and Galactic-potential scope}

For source-level diagnostics, we use the Monte Carlo mean and standard
deviation adopted in the main analysis. The source-level precision ratio is
defined as
\begin{equation}
 {\cal S}_{i}=
 \frac{|\langle a^{\rm Gal}_{i}\rangle_{\rm MC}|}
      {{\rm std}_{\rm MC}(a^{\rm tim}_{i})} .
 \label{eq:Ssource}
\end{equation}
We call a source high precision for this diagnostic when
${\cal S}_{i}>1$. This condition is satisfied by 7/26, 6/26, and 7/26
systems for MW2014, Gala2022, and McMillan2017, respectively.

\begin{figure}[t]
  \centering
  \includegraphics[width=0.96\columnwidth]
  {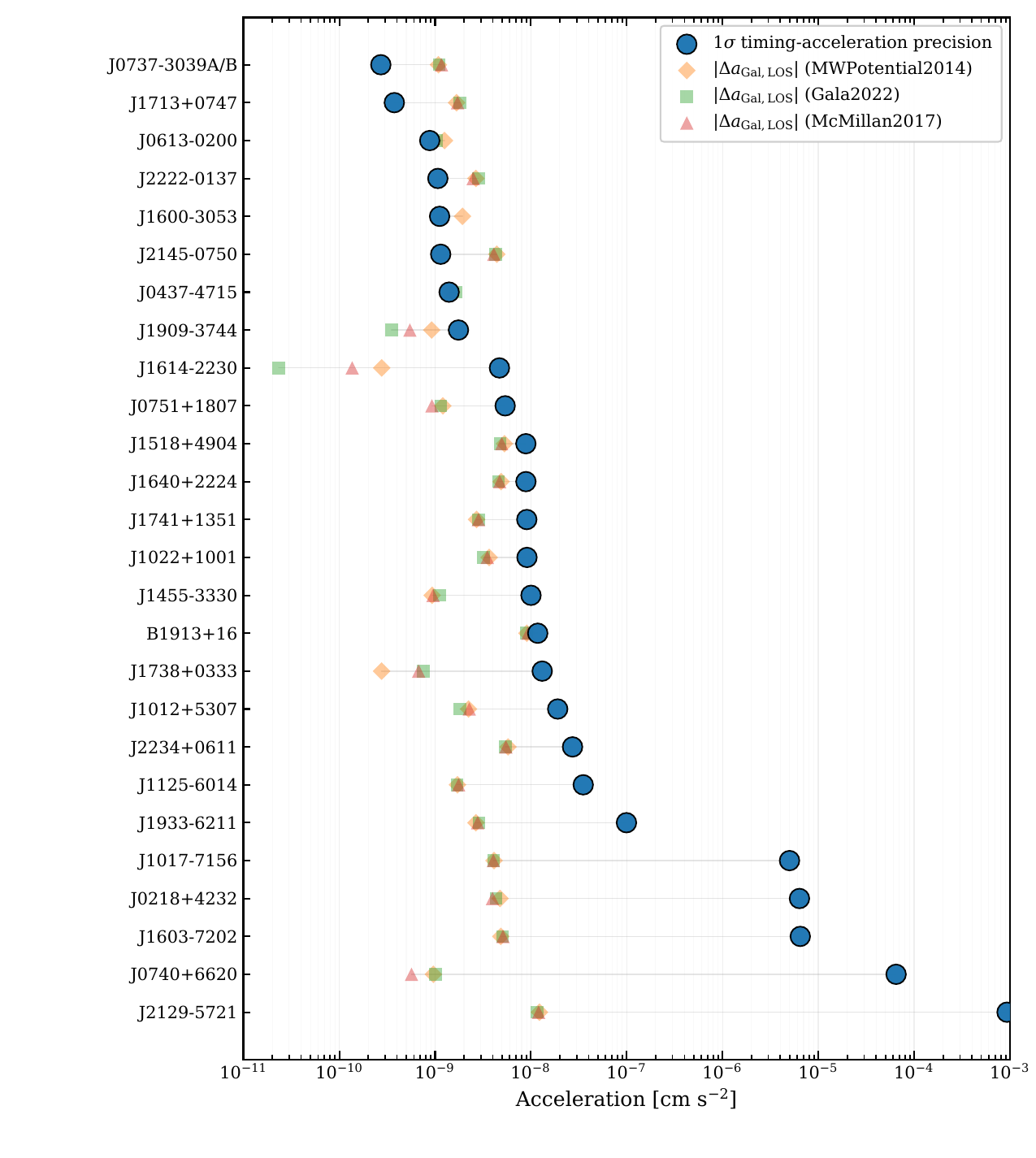}
  \caption{\textbf{Source-by-source acceleration scales.}
  The 26 binaries are ordered by timing-acceleration precision.
  Large circles show ${\rm std}_{\rm MC}(a^{\rm tim}_{i})$; smaller symbols
  show $|\langle a^{\rm Gal}_{i}\rangle_{\rm MC}|$ for MW2014, Gala2022, and
  McMillan2017.}
  \label{fig:precision}
\end{figure}

Because individual residual distributions can be asymmetric, we evaluate
source-level deviations using the shape-independent counting prescription of
Wang \textit{et al.} \cite{Wang2026}. For residual samples
$R^{(k)}_{i}$,
\begin{equation}
\begin{aligned}
 p_{+,i}
 &=\frac{N(R^{(k)}_{i}>0)}{N_{\rm MC}},\\
 p_{{\rm 2s},i}
 &=2\min[p_{+,i},1-p_{+,i}],
\end{aligned}
\label{eq:wang_p}
\end{equation}
with Gaussian-equivalent magnitude
\begin{equation}
 Z^{\rm eq}_{i}
 =\sqrt{2}\operatorname{erf}^{-1}(1-p_{{\rm 2s},i}),
\label{eq:wang_sigma}
\end{equation}
and the sign of $\langle R_i\rangle_{\rm MC}$ attached when quoting a signed
deviation. Among the high-precision sources, J0613$-$0200 is the only
system above $2\sigma$ for the two fixed potentials, at $+2.01\sigma$ for
MW2014 and $+2.25\sigma$ for Gala2022; no source reaches $3\sigma$.


Fig.~\ref{fig:precision} also motivates the adopted Galactic-potential
treatment. Since ${\cal S}_{i}>1$ for only 6--7 of the 26 binaries,
depending on the potential, the timing uncertainty exceeds the modeled
SSB--pulsar differential Galactic acceleration for most sources. A flexible
Galactic field is spatially varying, whereas the signal of interest is a
geometry-fixed common observer dipole. With the present heterogeneous sample,
jointly fitting both would therefore be strongly parameterization dependent.

We instead carry three published smooth Galactic potentials as separate
baselines and test for a coherent excess common acceleration. Their consistent vector fits demonstrate the stability of this low-dimensional measurement. A larger and more uniform
sample with full timing-solution covariances should eventually permit a joint
fit of Galactic-field structure and the observer dipole.





\section{Mass-reach calculation}
\label{app:numerics}

The fixed-mass heliocentric reaches for the full 26-source sample are
summarized in Table~\ref{tab:reach}. We quote the reach in AU for
$M<1\,\Msun$ and in pc for $M\geq1\,\Msun$.

The production residual ensemble contains $10^6$ joint Monte Carlo
realizations. The source-space calculation uses a dense radial grid from
100 AU to 12 kpc and an approximately uniform all-sky tessellation, with
hierarchical angular refinement near percentile boundaries,
high-information pulsar directions, and least-favorable regions.

\begin{table}[t]
\caption{\label{tab:reach}
Heliocentric reach for a compact object of fixed mass.
$D_p$ denotes the distance excluded over $p\%$ of sampled sky directions,
with $D_{100}$ corresponding to the least-favorable direction.
Distances are given in AU for $M<1\,\Msun$ and in pc for
$M\geq1\,\Msun$.}
\begin{ruledtabular}
\begin{tabular}{ccccc}
$M/\Msun$ & $D_{25}$ & $D_{50}$ & $D_{75}$ & $D_{100}$\\
\hline
$10^{-4}$ & 301.2 AU  & 255.1 AU  & 221.7 AU  & 136.3 AU \\
$10^{-3}$ & 952.3 AU  & 806.8 AU  & 701.0 AU  & 431.1 AU \\
$10^{-2}$ & 3011.5 AU & 2551.4 AU & 2216.8 AU & 1363.3 AU\\
$10^{-1}$ & 9523.3 AU & 8068.3 AU & 7010.0 AU & 4311.1 AU\\
$1$        & 0.146 pc & 0.124 pc & 0.107 pc & 0.066 pc \\
$10^{1}$   & 0.461 pc & 0.391 pc & 0.338 pc & 0.209 pc \\
$10^{2}$   & 1.465 pc & 1.239 pc & 1.068 pc & 0.661 pc \\
$10^{3}$   & 4.608 pc & 3.913 pc & 3.382 pc & 2.090 pc \\
$10^{4}$   & 14.587 pc & 12.353 pc & 10.689 pc & 6.608 pc \\
\end{tabular}
\end{ruledtabular}
\end{table}

\end{document}